# Evidence of intertwined pair density and charge density wave orders in $UTe_2$

Zhen Zhu[1], Yudi Huang[1], Julian May-Mann[2], Kaiming Liu[1], Zheyu Wu[3], Shanta R. Saha[4], Johnpierre Paglione[4,5], Alexander G. Eaton[3], Andrej Cabala[6], Michal Vališka[6], Eduardo Fradkin[1,7], and Vidya Madhavan[1,5]*

1. Department of Physics and Materials Research Laboratory, Grainger College of Engineering, University of Illinois at Urbana-Champaign, Urbana, IL, USA.
2. Department of Physics, Stanford University, Stanford, CA 94305, USA.
3. Cavendish Laboratory, University of Cambridge, Cambridge CB3 0HE, United Kingdom.
4. Maryland Quantum Materials Center, Department of Physics, University of Maryland, College Park, MD, USA.
5. Canadian Institute for Advanced Research, Toronto, Ontario, Canada.
6. Charles University, Faculty of Mathematics and Physics, Department of Condensed Matter Physics, Ke Karlovu 5, Prague 2, 121 16, Czech Republic.
7. Institute for Condensed Matter Theory, University of Illinois, Urbana, IL, USA.

* vm1@illinois.edu

**Abstract:**

The strongly correlated spin-triplet superconductor $UTe_2$ hosts an unusual landscape of magnetic-field-sensitive charge density wave (CDW) phases, positioning it as a compelling system for studying intertwined electronic orders. A central challenge is determining whether the observed charge modulations arise from a triplet pair density wave (PDW) order and, if so, how the anisotropic magnetic field response of triplet superconductivity is manifested in the CDW response. Here, using a scanning tunneling microscope equipped with a vector magnetic field, we systematically investigate the evolution and interrelation of distinct CDW orders. Complementing the previously identified incommensurate CDW peaks ($q_{i=1,2,3}$), we resolve an additional set of nondispersive modulations ($p_{i=1,2,3}$ and $h_{1,2}$) with distinct temperature and magnetic field dependencies. The $p_i$ CDW peaks vanish near $T_c$, while the $q_i$ peaks survive well above $T_c$ but are progressively suppressed by magnetic field in an anisotropic manner. The critical fields of the $q_i$ peaks mirror the directional hierarchy of $H_{c2}$, which suggests a PDW is present above the bulk $T_c$. This is consistent with a Landau free-energy picture where PDWs with wavevectors $p_i$ form above the bulk $T_c$, leading to composite CDW orders with wavevector $q_i$. Below $T_c$, the coupling of PDWs and uniform superconductivity leads to the $p_i$ CDWs. Together, these findings establish $UTe_2$ as a rare platform where both the parent PDW and descendant orders are directly resolved, enabling access to both the fundamental and emergent manifestations of PDW physics.

**Main text:**

Intertwined electronic orders represent a central theme in correlated quantum materials, where multiple orders like electronic nematicity, superconductivity, charge density wave (CDW), and spin density wave (SDW) do not merely coexist or compete but arise from a common microscopic origin[1-4]. Unlike conventional competition or coexistence between different phases, intertwined states often emerge in the overlapping regions of the phase diagram, exhibit comparable onset temperatures, and are linked by mutual coupling and symmetry constraints. A prominent manifestation of this paradigm is the pair density wave (PDW) state, a superconducting phase with a spatially modulated order parameter, which intertwines superconductivity and CDW order.

Accumulating experimental evidence supports the presence of PDW states in a variety of correlated systems, including cuprates[5-7], transition metal dichalcogenides[8-10], kagome metals[11,12], and iron-based superconductors[13,14]. In particular, the heavy-fermion compound $UTe_2$ has emerged as a promising platform for realizing this exotic order. $UTe_2$ is a spin-triplet superconductor with an unconventional pairing mechanism and an extremely large and anisotropic upper critical field ($H_{c2}^{b} > H_{c2}^{c} > H_{c2}^{a}$), reaching well beyond the Pauli limit[15]. Early-stage $UTe_2$ crystals exhibit superconductivity with $T_c \approx 1.6$K, field-reinforced superconductivity beyond 30 T[16], and signature of chiral spin-triplet pairing[17]. Next-generation crystals grown by molten flux methods exhibit an enhanced $T_c \approx 2.1$K[18-20]. Benefiting from their ultrahigh residual resistivity ratios and exceptional crystallinity, these samples provide an ideal platform that is less susceptible to extrinsic effects and point to a strengthened spin-triplet pairing component, as evidenced by higher $T_c$, increased $H_{c2}$ along all three crystallographic axes, and an expanded high-field superconducting phase[21].

An unusual incommensurate CDW order ($q_i$, $i$ =1, 2, 3) was observed in the early-stage $UTe_2$ samples, which melts near $H_{c2}$, suggesting that the observed charge modulations arise from PDWs[22-25]. In this picture, spin-triplet PDWs with momentum $q_i$, together with uniform spin-triplet superconductivity, naturally produce CDWs at momentum $q_i$. This framework accounts for the magnetic-field sensitivity of the CDWs through vortex-induced dislocations. However, several puzzles remain. For instance, if the CDWs are generated by the combination of PDWs and uniform superconductivity, they would be expected to vanish once uniform superconductivity is lost. Yet the field-sensitive CDWs persist well above $T_c$ up to ~5K.

Recent studies have offered new insights, but the interpretation of the data has varied across the community. The common feature between groups is the presence of spectral signatures with reduced wavevectors, distinct from the previously known CDW order[26-28] (here labeled as $h_1$, $h_2$, $p_1$, $p_2$, $p_3$). Peaks at $p_1$, $p_2$ and $p_3$ (henceforth referred to as $p_i$) were observed in applied magnetic fields at 30mK temperatures and attributed to primitive CDW components[26] owing to their commensurate relationship with $q_1$, $q_2$ and $q_3$, as each exhibits a wavevector of half the corresponding $q_i$. On the other hand in different studies, peaks at $h_1$, $h_2$ and $p_3$ were seen at energies within the superconducting gap and attributed to quasiparticle interference (QPI) from in-gap states along the nodal direction[27,28]. The microscopic origin of these multiple momentum space signatures and their relationship to the bulk homogeneous superconducting state is thus an important open question which has direct consequences for our understanding of the superconducting state.

In this work, we address these questions by studying next-generation $UTe_2$ samples ($T_c \approx$ 2.1K, Supplementary Fig. 1) using a vector-magnetic-field scanning tunneling microscope (STM). With this capability, we systematically investigate the temperature and field evolution of the established and newly emergent CDW orders in $UTe_2$. The $h_{1,2}$ and $p_3$ vectors are intrinsic to the zero-field state, whereas $p_1$ and $p_2$ are induced by applied magnetic fields. Energy dependent *dI/dV* maps establish that the $h_{1,2}$ and $p_3$ peaks, with wavevectors comparable to previous studies[27,28], remain nondispersive at energies well above the superconducting gap, confirming their classification as CDW modulations. We further show that the zero-field $p_3$ and field-induced $p_{1,2}$ peaks possess wavevectors that are half of $q_3$ and $q_{1,2}$. While this wavevector relationship may suggest that $q_i$ is a harmonic of $p_i$[26], the differing onset temperatures are incompatible with a simple harmonic interpretation. Notably, the $p_2$ and $p_3$ peaks are strongly coupled to superconductivity and vanish near the bulk $T_c$, while the $q_i$ peaks survive to higher temperatures. While the $p_1$ peak temperature dependence was not tracked, it is related by symmetry to the other peaks and is expected to show a similar temperature dependence as $p_2$ and $p_3$.

Our vector magnetic field response of the peaks conducted both below and above the superconducting $T_c$ uncover additional clues to their origin. Taken together and in combination with Landau free energy, our new data provide a refined explanation for both the newly observed and previously identified CDW peaks in terms of spin-triplet PDWs with wavevectors $p_i$ and uniform spin triplet superconductivity. In the new explanation, the PDWs have wavevectors $p_i$. The $q_i$ CDWs arise purely from the PDWs, and the $p_i$ CDWs arise from the combination of PDW and uniform superconductivity[a]. This explains why the $p_i$ peaks only show up in the STM data below the bulk $T_c$, and accounts for the hierarchical magnetic field destruction of the $q_i$ CDWs at 4.2K. The spin-triplet nature of the PDW and uniform superconductivity can also account for the fact that $p_1$ and $p_2$ are induced by a magnetic field.

$UTe_2$ crystallizes in an orthorhombic structure (space group Immm), in which Te atoms form chains along the crystallographic *b* axis, while U dimers oriented along the *c* axis assemble into chains extending along the *a* axis (Fig. 1a and b). STM measurement on a cleaved surface of $UTe_2$ reveals quasi-one-dimensional Te chains along the *a* axis (Fig. 1c), consistent with previous studies[17,23]. Each U atom is coordinated by two inequivalent tellurium atoms, Te1 (light blue) and Te2 (dark blue). Owing to the centrosymmetric nature of the bulk structure, $UTe_2$ can cleave along either the (01-1) or (011) surface[17,23], leading to surface terminations with Te1 and Te2 chains swapping their relative positions. On both cleavage surfaces, the Te1 chains appear slightly elevated compared to the Te2 chains, serving as a distinguishing feature[17]. Based on the spatial arrangement of Te1 and Te2 atoms observed in the high-resolution topography (inset of Fig. 1c) and following established criteria, we tentatively identify the cleavage plane in this study as (01-1). This identification in turn allows us to establish the correspondence between the applied magnetic field directions and the crystallographic axes in the subsequent analysis.

---

[a] In the previous definition[23], the PDWs had wavevector $q_i$ and the $q_i$ CDWs were taken to arise from the combination of PDWs and uniform superconductivity.

The atomically resolved image acquired on a defect-free area below $T_c$ reveals chains of Te1 atoms along *a*-axis (Fig. 1c), and the corresponding fast Fourier transform (FFT) is shown in Fig. 1d. Beyond the previously reported $q_i$ CDW orders (marked by squares), a new set of $h_{1,2}$ and $p_3$ peaks emerge at zero field (marked by blue/red triangles and a green circle). By performing spectroscopic imaging via *dI/dV* maps, we obtain the real-space distribution of the local density of states (LDOS), whose FFTs capture the energy-dependent spectral weight in momentum space, enabling a distinction between quasiparticle interference and CDW signatures. Figures 1e-g present the FFTs of *dI/dV* maps at the indicated energies, including one within and two well above the superconducting gap, where the $h_{1,2}$ and $p_3$ peaks are clearly visible (see Supplementary Figs. 2 and 3 for the full dataset). We investigate the energy dependence of the $h_{1,2}$ and $p_3$ peaks, by performing line cuts along the three momentum directions (Fig.1e). These peaks remain nondispersive across the measured energy range extending to energies far above the superconducting gap (Fig.1 i-k), manifesting their origin as CDW orders rather than their previous identification as QPI features[27,28]. The wavevectors of $h_{1,2}$ and $p_3$ orders are $\{\boldsymbol{h_1}, \boldsymbol{h_2}, \boldsymbol{p_3}\} = \{(-0.26, 0.14), (0.26, -0.14), (0, -0.28)\}|q_{Te}|$, where $|q_{Te}|$ denotes the Bragg wavevector of the Te lattice. All three vectors have comparable magnitudes and together form a hexagonal lattice–like periodic modulation in real space (Supplementary Fig. 4). Interestingly, the line profile extracted along line 3 in Fig. 1e shows that the magnitude of the $p_3$ peak is approximately half that of $q_3$ (Fig. 1h). The significance of this relationship will be discussed later in the paper. It is also worth noting that the interference of defects most likely hindered the observation of the $h_{1,2}$ and $p_3$ CDW orders in previous studies[22-24]. This is because in reciprocal space, defect-induced electronic states can generate intense spectral weight near zero momentum, with momentum components similar to those of the $h_{1,2}$ and $p_3$ wavevectors, thereby smearing or obscuring these intrinsic CDW features (Supplementary Fig.5).

A systematic, temperature-dependent study was carried out to determine the onset temperatures of the newly identified $h_{1,2}$ and $p_3$ orders. Since the CDW transition in $UTe_2$ currently lacks a clear bulk signature[29-31], STM enables a direct characterization of the surface CDW evolution with temperature in a momentum-resolved manner. Figures 2a–g present a series of normalized FFTs of STM topographies acquired at the indicated temperatures. Across the superconducting transition at $T_c$= 2.1K, the $p_3$ CDW peak exhibits the most significant change among all CDW components, being strongly suppressed above the transition temperature. This is clearly reflected in Figs. 2h and i, which show the $h_{1,2}$, $p_3$ and $q_i$ peak intensities extracted from the FFTs. The $q_i$ CDW intensities decrease with temperature and are markedly reduced at 4.8 K, consistent with our previous report[23]. In contrast, the $p_3$ CDW intensity decreases progressively with increasing temperature, becoming close to the background above $T_c$. This suggests a relationship between $p_3$ and the uniform superconducting order. Equally importantly, the $p_3$ and $q_3$ peaks emerge at different onset temperatures, which strongly indicates that they are not harmonically related wavevectors as was previously hypothesized[26], and instead points to a different mechanism governing the interplay between the $p_i$ and $q_i$ peaks. We will investigate this possibility in the remainder of the paper.

Before proceeding with further discussions on the possible origins of the $p_3$ peak, we look for signatures of $p_1$ and $p_2$ peaks which were previously observed in magnetic fields at 30mK temperatures[26]. To this end, we employ a vector magnet that enables independent tuning of all three magnetic field components, with their orientations relative to the crystallographic axes of $UTe_2$ defined throughout this work as illustrated in Fig. 1b. Specifically, the laboratory axis $B_y$ is aligned with crystallographic *a*-axis, while $B_x$ and $B_z$ are tilted counterclockwise by approximately 12° from the *c*- and *b*-axes, respectively (see Supplementary Fig. 6 for details). For simplicity, we label the magnetic field directions $B_x$, $B_y$, $B_z$, as $B_{c^*}$, $B_a$, $B_{b^*}$ to reflect their approximate alignment with the crystallographic *c*-, *a*-, and *b*-axes, respectively. This field-axis convention is based on the assumption that the sample cleaved along the (01-1) plane. We note that density functional theory calculations indicate that Te2 p orbitals dominate the density of states at the Fermi level[32], suggesting that the visually dominant chains in the STM topography may correspond to Te2 atoms, implying a cleavage along the (011) plane (Supplementary Fig. 6). This would modify the relative orientation between the laboratory and crystallographic axes, but $B_{b^*}$ and $B_{c^*}$ remain primarily aligned with the *b*- and *c*-axes, leaving our conclusions unchanged.

Figures 3a–d present the FFTs of topographic images acquired at 300 mK under the indicated magnetic field directions and magnitudes. Compared with the zero-field result (Fig. 3a), a consistent feature across field configurations in Figs. 3b-d is the emergence of a field-induced $p_2$ peak, whose wavevector magnitude is half that of $q_2$. Energy-dependent measurements further confirm that the $p_2$ peak, originates from a CDW order (see Supplementary Fig. 7 for details). The CDW peaks observed at 0 T ($h_1$, $h_2$, $p_3$, $q_1$, $q_2$, $q_3$) and those appearing under finite magnetic fields ($p_1$ and $p_2$) are summarized in Fig. 3e, which illustrates their distribution in momentum space. The analogous wavevector relations, $p_{2,3} = 1/2q_{2,3}$, suggest that the $p_2$ and $p_3$ CDWs share a common origin. To explore this connection, we perform systematic vector field- and temperature- dependent measurements, focusing on the evolution of the $p_2$ order.

We first construct the field-temperature phase diagram with magnetic field along $b^*$ ($B_{b^*}$) by plotting the intensity of the $p_2$ peak. As seen in Fig. 3f (see full dataset in Supplementary Fig. 8), at 300 mK, the $p_2$ peak appears at an onset field $H_{onset}$ of ~5T and persists up to 8T. As the temperature increases to 1.6 K, the field range sustaining $p_2$ shrinks to approximately 5–6 T, and its intensity becomes weaker. The field at which the intensity is suppressed below the background ($H_{sup}$) decreases with increasing temperature, and the $p_2$ feature disappears entirely when the temperature exceeds $T_c$ (i.e., 2.5K). Similar measurements under $B_a$ and $B_{c^*}$ show a comparable trend (Figs. 3g and h). For a given $B_a$ magnetic field, the $p_2$ peak progressively weakens with increasing temperature and disappears above $T_c$ (Fig. 3g, see full dataset in Supplementary Fig. 9). At 300 mK, the $p_2$ peak strengthens with increasing $B_{c^*}$ up to 2 T (Fig. 3h, see full dataset in Supplementary Fig. 10). These results indicate that the field-induced $p_2$ modulation exhibits a similar temperature dependence to the zero-field $p_3$ peak and is connected to uniform superconductivity. $H_{onset}$ for $p_2$ is substantially larger for fields along $b^*$ than along $a$ and $c^*$ (~ 0.5 T at 300mK), indicating a higher threshold for the emergence of $p_2$ near the $b$ axis, consistent with the $b$-axis being the magnetic hard axis of $UTe_2$ and hosting the largest $H_{c2}$. Additionally, the $p_1$ CDW peak, which is the mirror counterpart of $p_2$, emerges under $B_{b^*}$ in another field of view (Supplementary Fig. 11). Finally, we note that in contrast to the $p_3$ and $p_2$ CDWs, the $q_i$ CDWs persists under the same field conditions at temperatures below 2.5 K.

So far, we have established: (i) that all $h_{1,2}$ and $p_{1,2,3}$ orders are visible at energies far above the superconducting gap and do not disperse which suggests that they are not related to nodal in-gap QPI; (ii) $p_3$ is visible in zero field while $p_{1,2}$ emerge at a finite field; and (iii) the differing magnetic field behavior and onset temperatures of the $q_i$ and $p_i$ orders indicate that despite their commensurate relationship i.e., $p_i = 1/2q_i$, the $q_i$ order is not a simple harmonic of the $p_i$ order.

The final piece of the puzzle is to ascertain whether the original $q_i$ CDW order exhibits field sensitivity at elevated temperatures (>$T_c$). To do this we performed vector-field measurements at 4.2 K. As expected from the temperature dependence shown in Fig. 2, the FFT of the LDOS map at 4.2K and zero field (Fig. 4a) reveal only the energy-independent $q_i$ peaks (Supplementary Fig. 12), with a schematic of these CDW wavevectors shown in Fig. 4b. Upon applying magnetic fields along or near the crystallographic axes ($B_{b^*}$ = 4.0 T, $B_a$ = 0.4 T, and $B_{c^*}$ = 1.5 T), the $q_i$ peaks exhibit a reduction in intensity but remain discernible (Figs. 4c, e and g). With further increase in field strength ($B_{b^*}$ = 8.0 T, $B_a$ = 0.8 T, and $B_{c^*}$ = 3.0 T), the $q_i$ peaks are nearly or completely suppressed along all directions (Figs. 4d, f and h). All datasets were acquired over the same spatial region using identical bias setpoints and energies, ensuring that the observed changes are exclusively field-induced. Compared to the results at lower temperatures where the $q_i$ modulation persists under the same field magnitudes (Supplementary Figs. 8-10), we conclude that $H_{sup}$ of the $q_i$ order decreases dramatically with increasing temperature. Specifically, at 4.2K, $H_{sup}$ falls within the ranges 4.0< $H_{sup}^{b^*}$<8.0 T, 0.4< $H_{sup}^{a}$<0.8 T and 1.5< $H_{sup}^{c^*}$<3.0 T along the $b^*$-, $a$-, and $c^*$-axis directions, respectively.

By extracting the normalized intensity of each $q_i$ peak from the FFTs, we can track its evolution as a function of magnetic field strength for each crystallographic direction (Figs. 4i–k). Scaling the applied field $B_n$ ($n$=$a$, $b^*$, $c^*$) by the corresponding upper critical field $H_{c2}$ reveals the anisotropic relation $H_{sup}^{b^*} > H_{sup}^{c^*} > H_{sup}^{a}$ for all $q_i$ peaks, in agreement with the established hierarchy of $H_{c2}^{b^*} > H_{c2}^{c^*} > H_{c2}^{a}$. Since superconductivity is fully suppressed at 4.2 K, this correlation between the anisotropy of $H_{sup}$ and $H_{c2}$ provides compelling evidence that the $q_i$ CDW order is intertwined with a PDW component.

Finally, with comprehensive temperature- and field-dependent data in hand, we are now positioned to address the origins of the $p_{1,2,3}$ ($p_i$) peaks and their relation to the previously identified $q_{1,2,3}$ ($q_i$) CDW peaks, as well as the emergence of the $h_1$ and $h_2$ peaks. In our previous work, the $q_i$ CDW peaks were identified as composite orders which formed from the combination of wavevector $q_i$ PDWs and uniform superconductivity[23]. With the new information in this study, we can provide a more refined explanation for the results.

As we shall now show, the magnetic field sensitivity of the $q_{1,2,3}$ CDW peaks above the bulk $T_c$ and the formation of the $p_3$ CDW at the bulk $T_c$, can both be explained by the existence of a PDW that forms with a higher critical temperature than the bulk $T_c$, and coexists with uniform superconductivity below $T_c$. Let us consider the following Landau free energy for the CDW and PDW orders:

$$
\begin{aligned}
\mathcal{F} = & \sum_i \alpha_i (T - T_{PDW}) |\boldsymbol{d}_{p_i}|^2 + \lambda_i |\boldsymbol{d}_{p_i}|^4 \\
& + \alpha_0 (T - T_c) |\boldsymbol{d}_0|^2 + \lambda_0 |\boldsymbol{d}_0|^4 \\
& + r_1 \sum_i |\rho_{p_i}|^2 + r_2 \sum_i |\rho_{2p_i}|^2 + r_4 [\rho_{2p_3-2p_2} + \rho_{2p_3-2p_1}] \\
& + g_1 \sum_i \rho_{p_i} [\boldsymbol{d}_0 \cdot \boldsymbol{d}_{p_i}^* + \boldsymbol{d}_0^* \cdot \boldsymbol{d}_{-p_i}] + g_2 \sum_i \rho_{2p_i} \boldsymbol{d}_{-p_i} \cdot \boldsymbol{d}_{p_i}^* \\
& + \gamma_1 \sum_i \rho_{p_i} \rho_{p_i} \rho_{2p_i}^* + \gamma_2 [\rho_{2p_3} \rho_{2p_2}^* \rho_{2p_3-2p_2}^* + \rho_{2p_3} \rho_{2p_1}^* \rho_{2p_3-2p_1}^*] + c.c.
\end{aligned}
\tag{1}
$$

Here, $\boldsymbol{d}_{p_i}$ denotes the $\boldsymbol{d}$-vector of the *triplet* PDW order parameter at wavevector $p_i$. For spin-1/2 fermions, the PDW gap function is a 2x2 matrix, and is related to the $\boldsymbol{d}$-vector as $\Delta_{p_i}(\boldsymbol{r}) = e^{i p_i \cdot \boldsymbol{r}} [\boldsymbol{d}_{p_i} \cdot \boldsymbol{\sigma}] i\sigma_y$, where $\boldsymbol{\sigma} = (\sigma_x, \sigma_y, \sigma_z)$ are the 2x2 Pauli matrices. Similarly, $\boldsymbol{d}_0$ is the $\boldsymbol{d}$-vector of the uniform triplet superconducting order parameter. $T_{PDW}$ is the transition temperatures of the PDW, which we take to be the same as the temperature where the $q_i$ CDW forms (approximately 5K). $T_c$ is the critical temperature for the bulk uniform superconductivity (approximately 2.1 K). Note that we have not included all symmetry allowed terms in the free energy. In particular, in Eq. (1) we have not included possible bi-quadratic terms, which will encode competitions or cooperation between different phase orderings. However, these additional terms are not necessary for discussing the PDW induced CDWs. Also, in Eq. (1) we have not included the effects of quenched random disorder (i.e., impurities) which couples strongly to any non-uniform order and, ultimately, destroys long range PDW and CDW orders over length scales larger than the field of view of our STM experiments.

Table 1 tabulates the observed CDW vectors and their connections to the PDW and uniform superconductivity (see Fig. 3e for their momentum-space distribution). In the following, we discuss the microscopic origin of these CDWs at different temperature regimes. When $T>T_{PDW}$, both PDW and superconducting orders are disordered ($\langle \boldsymbol{d}_{p_i} \rangle = \langle \boldsymbol{d}_0 \rangle = 0$), thus no CDW signal is observed. In the intermediate regime $T_{PDW}>T>T_c$, the PDWs become ordered ($\langle \boldsymbol{d}_{p_i} \rangle \neq 0$) while the uniform superconductivity remains disordered ($\langle \boldsymbol{d}_0 \rangle = 0$). The PDWs give rise to charge modulations at wavevector $2p_i$, $\rho_{q_i} \propto \boldsymbol{d}_{p_i} \cdot \boldsymbol{d}^*_{p_i}$, due to the tri-linear coupling between the different order parameters. These modulations correspond to the experimentally observed $q_i$ CDW peaks. The magnetic-field sensitivity of the $q_i$ peaks in this regime is governed by the same mechanism proposed in our previous work[23].

At temperatures below $T_c$, both the PDW and uniform superconductivity orders are established ($\langle \boldsymbol{d}_{p_i} \rangle \neq 0, \langle \boldsymbol{d}_0 \rangle \neq 0$). The combination of PDWs and uniform superconductivity can give rise to wavevectors $p_i$ CDWs, $\rho_{p_i} \propto \boldsymbol{d}_{p_i} \cdot \boldsymbol{d}^*_0 + \boldsymbol{d}^*_{-p_i} \cdot \boldsymbol{d}_0$. Among these CDWs, only $p_3$ is observed at zero field, while $p_1$ and $p_2$ emerge under an applied magnetic field and are theoretically expected to arise from the same superconductivity-coupled mechanism. Within this theory, it implies that $\boldsymbol{d}_0$ and $\boldsymbol{d}^*_{\pm p_{1,2}}$ are orthogonal to each other in zero magnetic field, while $\boldsymbol{d}_0$ and $\boldsymbol{d}^*_{\pm p_3}$ are not. The occurrence of the $p_1$ and $p_2$ CDWs in finite magnetic field, is explained by the fact that an external magnetic field can rotate $\boldsymbol{d}$-vectors and cause them to align. It is worth noting that if $\boldsymbol{d}_0$ and $\boldsymbol{d}^*_{-p_{1,2}}$ are actually orthogonal in zero magnetic field, there will be composite SDW order with wavevector $p_{1,2}$, $S_{p_i} \propto \boldsymbol{d}_0 \times \boldsymbol{d}^*_{p_i}$. As we shall discuss in the next section, the PDW would have to be localized at the surface of the system in a consistent theory, and the SDWs would be localized at the surface as well. Detecting the SDWs would require a surface-sensitive probe, such as spin-polarized STM.

From this analysis, we find that all the $q_i$ and $p_i$ CDWs orders can be consistently understood as arising from PDWs and uniform superconductivity. This scenario does, however, rest on the strong assumption that the PDWs form above the bulk $T_c$. Given that there is no bulk evidence of a PDW above $T_c$, such a state would have to be localized to the surface of the sample. In principle, a surface PDW can occur above the bulk $T_c$, as the energetics of the surface may be very different than those of the bulk. A microscopic theory would be needed to determine if a surface PDW can, in fact, arise in this way.

Based on the refined PDW framework, we are now able to elucidate the origin of the $h_1$ and $h_2$ modulations. In fact, the $h_1$ and $h_2$ modulations can be interpreted as harmonic-derived, higher-order composite CDWs. Since $h_1 = 2p_3 - 2p_2 = q_3 - q_2$, the equation of motion for $\rho_{2p_3-2p_2}$ show that the $h_1$ CDW can arise as a composite of the $q_3$ and $-q_2$ CDWs, or equivalently, as a fourth-order composite of PDW components. Similarly, the $h_2$ CDW corresponds to $\rho_{2p_3-2p_1}$ and can arise as a composite of the $q_3$ and $-q_1$ CDWs. This explanation also accounts for the similarities between the magnetic field responses of the $h_{1,2}$ CDWs (Supplementary Figs. 8 and 9) and those of the $q_i$ CDWs[23].

Our results reveal a previously unrecognized hierarchy of intertwined electronic orders in $UTe_2$. The vector-field and temperature-dependent measurements of the high-temperature $q_{1,2,3}$ and low-temperature $h_{1,2}$, $p_{1,2,3}$ CDW modulations point towards a PDW-based explanation, where wavevector $p_i$ PDWs form above $T_c$, leading to the $q_{1,2,3}$ and $h_{1,2}$ charge modulations, while the $p_{1,2,3}$ peaks emerge through coupling to uniform superconductivity below $T_c$. This is captured by a Landau free energy which couples the PDW, uniform superconductivity and CDW order parameters to one another. Our findings offer crucial insights into the nature of PDW states and their role in generating intertwined orders in strongly correlated systems.

## Methods

Single crystals of $UTe_2$ grown by a molten flux method were used. The growth and characterization are mentioned in detail elsewhere[19]. The STM measurement were performed in a Unisoku STM operating under ultra-high vacuum and a vector magnetic field ($H_x$:4T, $H_y$:1T, $H_z$:9T). The $UTe_2$ single crystals were cleave in situ at about 90K and then transferred immediately into STM head. All the measurements were carried out at 300 mK (unless otherwise specified) using tungsten tips treated with heating in situ. *dI/dV* signals were acquired by a standard lock-in amplifier with modulation of 500 μV at 991 Hz, unless otherwise noted.

**Acknowledgments:** STM studies at the University of Illinois, Urbana-Champaign were supported by the US Department of Energy, Office of Science, Office of Basic Energy Sciences, Materials Sciences and Engineering Division under award number DE-SC0022101. V.M. acknowledges partial support from Gordon and Betty More Foundation's EPiQS Initiative through grant GBMF4860 and the Quantum Materials Program at CIFAR where she is a Fellow. Theoretical work was supported in part by National Science Foundation grant DMR 2225920 (E.F.). A.G.E. acknowledges support from the Henry Royce Institute for Advanced Materials through the Equipment Access Scheme enabling access to the Advanced Materials Characterization Suite at Cambridge, grant numbers EP/P024947/1, EP/M000524/1 & EP/R00661X/1; and from Sidney Sussex College (University of Cambridge). M.V and A.C. acknowledge support by the Czech Science Foundation GAČR under the Junior Star Grant No. 26-21795M (STiUS). Crystals were grown and characterized in MGML (mgml.eu), which is supported within the program of Czech Research Infrastructures (project No. LM2023065). Research at the University of Maryland was supported by the Gordon and Betty Moore Foundation's EPiQS Initiative Grant No. GBMF9071, the NIST Center for Neutron Research, and the Maryland Quantum Materials Center.

**Author contributions:** Z.Z. and V.M. conceived the experiments. Z.W., S.R.S., J.P., A.G.E., A.C. and M.V. provided the characterized single crystals. Z.Z., Y.H. and K.L. obtained the STM data. Z.Z., Y.H. and V.M. carried out the analysis and J.M.-M. and E.F. provided the theoretical input on the interpretation of the data. Z.Z., V.M., J.M.-M. and E.F. wrote the paper with input from all authors.

**Competing interests:** The authors declare no competing interest.

**Figures:**

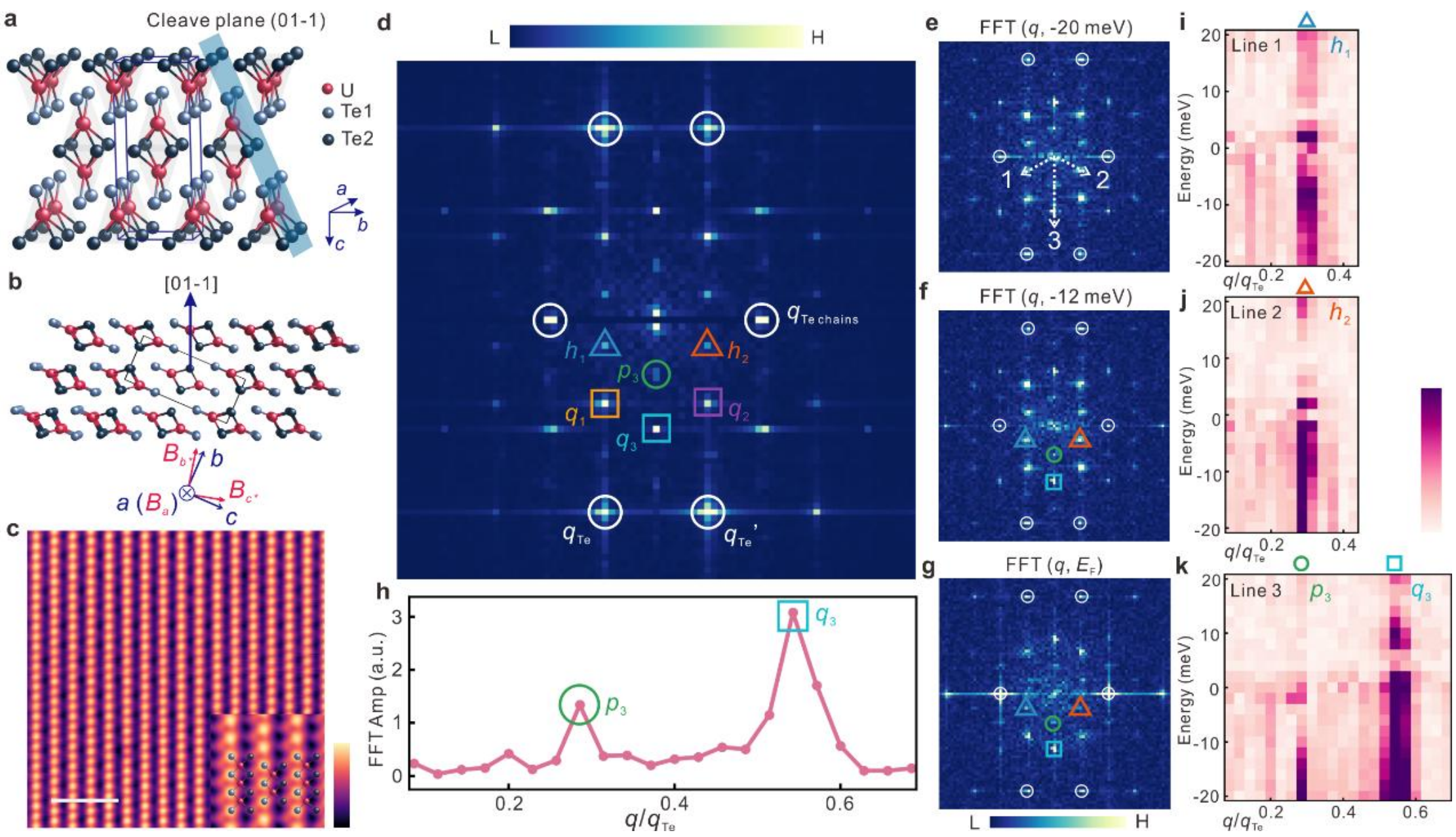

**Fig.1 | Crystal structure and two distinct sets of charge density wave orders of $UTe_2$.** **a**, Structural framework of $UTe_2$. The orthorhombic unit cell and the (01-1) cleavage plane are delineated by a rectangular box and a blue plane, respectively. Each U (red) atom is coordinated by two inequivalent Te sites, denoted Te1 (light blue) and Te2 (dark blue). **b**, Side-view schematic of (01-1) plane. Relations between externally applied magnetic fields ($B_a$, $B_{b^*}$, and $B_{c^*}$) and the crystallographic axes are illustrated. The $B_a$ direction is parallel to the *a*-axis, while the $B_{b^*}$ and $B_{c^*}$ fields are counterclockwise tilted by approximately 12° relative to the crystallographic *b* and *c* axes, respectively. **c**, Atomic-resolution image acquired on defect free region at 300 mK ($V$=20 mV, $I$=100 pA). Scale bar, 3nm. The inset shows the configuration of Te1 (light blue) and Te2 (dark blue) atoms with lattice schematic overlaid on top. **d**, Fourier transform of the topography shown in **c**. The Bragg peaks arising from Te1 atoms are marked by white circles ($\pm q_{\text{Te}}$, $\pm q_{\text{Te}}'$, $\pm q_{\text{Te chains}}$; only one side is labeled for clarity). Squares colored yellow ($q_1$), purple ($q_2$) and cyan ($q_3$) denote the experimentally confirmed set of CDW peaks. Additionally, a newly observed set of CDW orders under zero field is indicated by blue ($h_1$) and red ($h_2$) triangles, as well as a green circle ($p_3$). **e**-**g**, FFTs of LDOS maps shown in Supplementary Fig. 2 at 300 mK for the indicated energies. The new set of CDW modulations ($h_1$, $h_2$ and $p_3$), marked by blue/red triangles and a green circle, is clearly visible at all energies. The Bragg peaks are marked by white circles ($V$=20 mV, $I$=150 pA). **h**, Line cut extracted along line 3 indicated in **e**. The magnitude of $p_3$ is approximately half that of $q_3$. **i**-**k**, The intensity maps plotted by line cuts extracted from FFTs at various energies along the dashed lines in **e**. The magnitude of the $h_1$, $h_2$ and $p_3$ wavevectors remains constant with energy, consistent with their identification as CDW signals. All three CDW wavevectors exhibit similar magnitudes, approximately 0.28 $q_{\text{Te}}$.

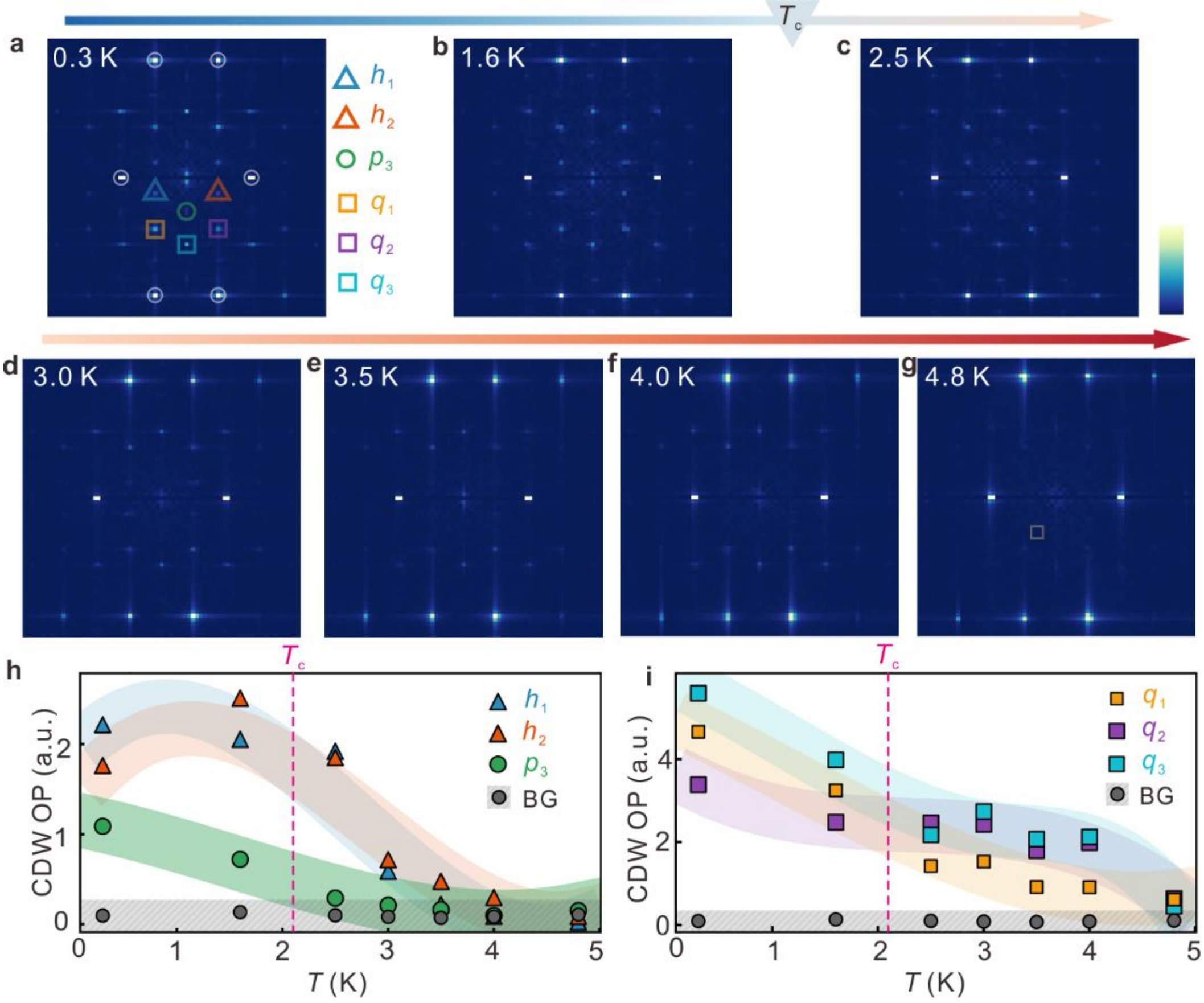

**Fig.2 | Temperature-dependent suppression of the $h_{1,2}$, $p_3$ and $q_i$ CDW orders. a-g.** FFTs acquired on the same area at the indicated temperatures, maintaining identical set points ($V$=20 mV, $I$=100 pA). The intensities of FFT images are normalized by $q_{Te}$ and $q_{Te}$'. The amplitude of the $p_3$ peak (green circle) rapidly diminishes upon crossing the superconducting transition temperature ($T_c$=2.1 K) of $UTe_2$. Peaks $h_1$ (blue triangle) and $h_2$ (red triangle) vanish at approximately 3.5 K. The intensity of the previously identified $q_i$ CDW continuously weakens with increasing temperature, becoming significantly attenuated by 4.8 K. The Bragg peaks are marked by white circles in **a**. **h,** Temperature dependence of the intensities of $h_{1,2}$, $p_3$ peaks. **i**, Same as **h**, but for the $q_i$ peak intensities. Extracted peak intensities are normalized relative to $q_{Te}$ and $q_{Te}$'. Scatter points are color-coded according to their corresponding CDW labels, and shaded areas of matching colors serve as visual guides to indicate overall intensity trends. The background intensity (BG), extracted from FFT regions free of CDW peaks (gray square in **g**), is shown as gray solid circles with hatched shading and remains nearly constant with temperature.

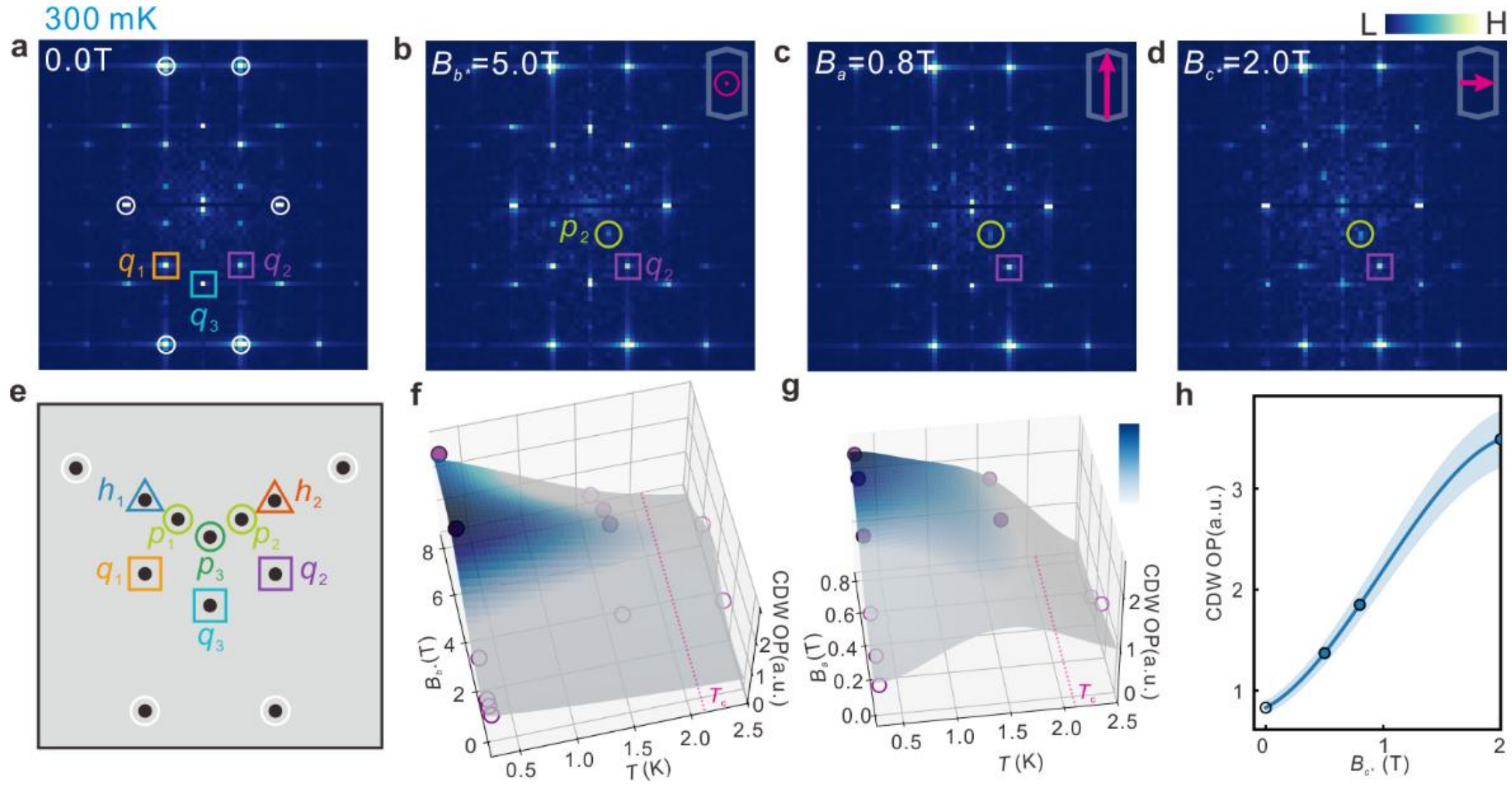

**Fig.3 | Magnetic field and temperature evolution of the field-induced CDW peak $p_2$.** **a-d,** FFTs obtained on the same area at the indicated magnetic fields under consistent experimental conditions ($T$=300 mK, $V$=20 mV, $I$=100 pA). The orientations of applied fields ($B_{b^*}$, $B_a$, and $B_{c^*}$) relative to crystal axes are illustrated in Fig. 1b. An additional CDW peak, $p_2$ (light green circle), emerges under various field configurations and exhibits a wavevector magnitude equal to half of $q_2$ (purple square). The Bragg peaks are marked by white circles in **a**. **e,** Schematic summary of CDW peaks observed at 0 T ($h_1$, $h_2$, $p_3$, $q_1$, $q_2$, $q_3$) and under finite magnetic fields ($p_1$ and $p_2$) at 300mK. White circles indicate Bragg points. **f,** Three-dimensional plot of the $p_2$ CDW peak intensity as functions of temperature and magnetic field along the $B_{b^*}$ direction. Peak $p_2$ appears at 5 T and persists at least to 8 T at 300 mK, remains detectable at narrower field range at 1.6 K, and is absent at 2.5 K. **g,** Same as **f,** but with magnetic field applied along the $B_a$ direction. The peak arises around 0.5 T at 300 mK, intensifies with increasing field, persists at 1.6 K under corresponding magnetic fields, and vanishes at 2.5 K. **h,** Dependence of $p_2$ CDW peak intensity on magnetic field applied along the $B_{c^*}$ direction at 300mK. The peak emerges at approximately 0.5 T and is enhanced by further increasing the field. Solid circles and open circles in **f-h** indicate the presence and absence of the $p_2$ peak at the corresponding magnetic fields and temperatures, respectively. All intensities presented in **f-h** are normalized with respect to $q_{Te}$ and $q_{Te}$'. Complete FFT datasets are provided in Supplementary Figs. 8-10.

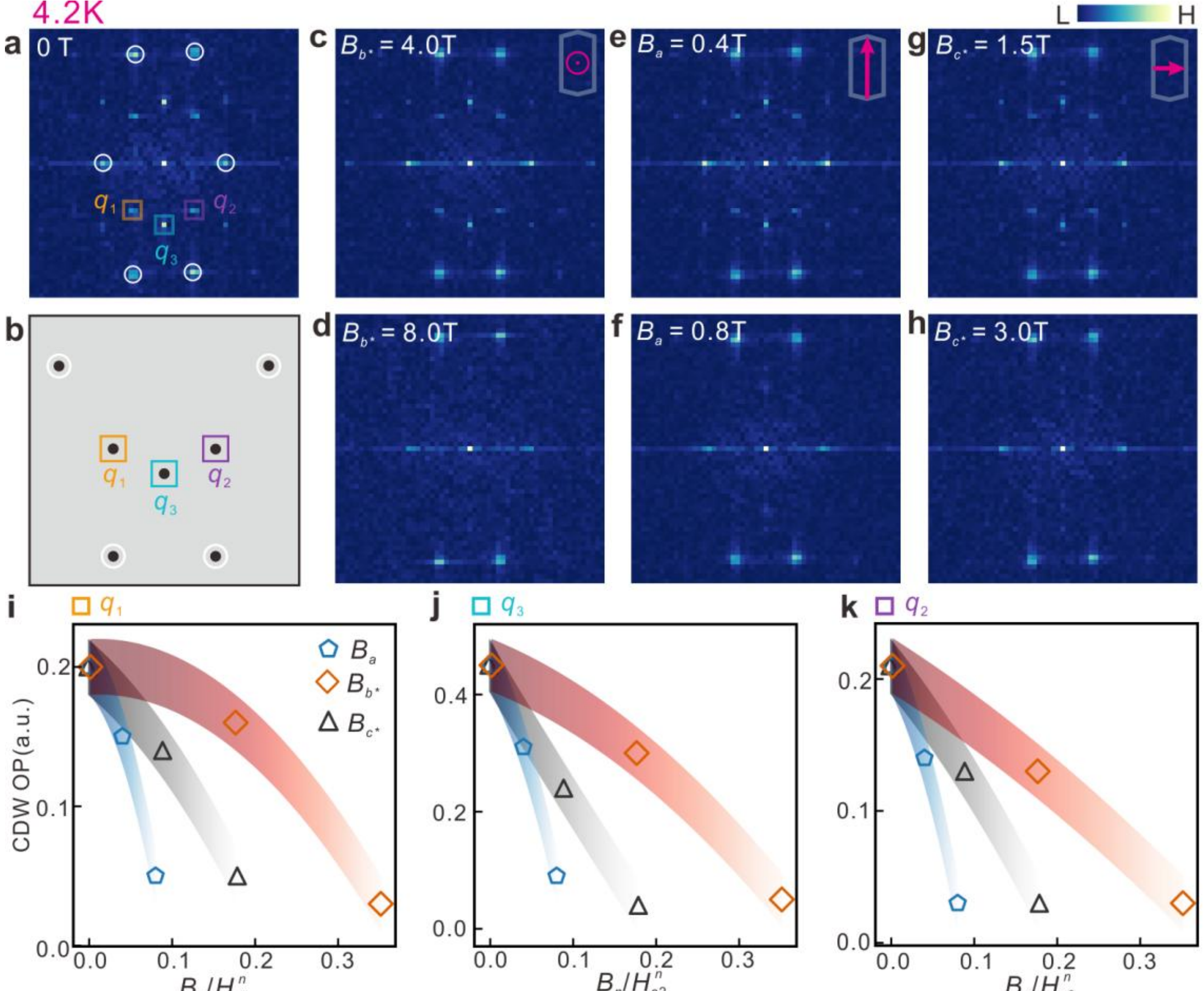

**Fig.4 | Magnetic field dependence of the $q_i$ CDW orders above $T_c$ (4.2K). a,** FFT of the LDOS map acquired at -20mV and 4.2K under zero magnetic field, where only the $q_i$ CDW is visible. The Bragg peaks are marked by white circles ($V$ = 20 mV, $I$ = 150 pA). **b**, Schematic of the $q_i$ CDW peaks observed at 4.2K. White circles indicate Bragg peaks. **c-h**, FFTs of LDOS maps acquired at –20 mV and 4.2 K under the indicated magnetic fields. All measurements were performed in the same area as in **a**, using identical set-point conditions ($V$ = 20 mV, $I$ = 150 pA). The $q_i$ CDW peaks are completely suppressed in **d** ($B_{b^*}$=8.0 T) and **h** ($B_{c^*}$=3.0 T). Peaks $q_1$ and $q_2$ are entirely suppressed, whereas $q_3$ remains only barely visible in **f** ($B_a$=0.8 T). The intensities of FFT images are normalized by $q_{\mathrm{Te}}$ and $q_{\mathrm{Te}}$'. **i-k,** Intensity variations of the three CDW peaks $q_1$ (yellow square), $q_3$ (cyan square), and $q_2$ (purple square) as functions of vector magnetic fields $B_n$ ($n$=$a$, $b$*, $c$*), respectively. The field values along each direction are scaled by the corresponding $H_{c2}^{n}$ ($n$=$a$, $b$*, $c$*) of $UTe_2$. All extracted intensities of the $q_i$ peaks are normalized with respect to $q_{\mathrm{Te}}$ and $q_{\mathrm{Te}}$'. The magnetic fields required to fully suppress the $q_i$ CDW peaks differ along the three field directions following the order $H_{sup}^{b^*} > H_{sup}^{c^*} > H_{sup}^{a}$, which aligns with the corresponding trend of $H_{c2}$, namely, $H_{c2}^{b^*} > H_{c2}^{c^*} > H_{c2}^{a}$.

**Table 1. Summary of the observed charge density wave (CDW) vectors and their relationship to the pair density wave (PDW) and uniform superconductivity.**

| CDW vectors | Category | Physical origin | Representative expressions |
|---|---|---|---|
| $p_i$ ($i$=1,2,3) | Fundamental (low $T$) | Composite of $p_i$ PDWs and uniform superconductivity | $\rho_{p_i} \propto \boldsymbol{d}_{p_i} \cdot \boldsymbol{d}_0^* + \boldsymbol{d}_{-p_i}^* \cdot \boldsymbol{d}_0$ |
| $q_i$ ($i$=1,2,3) | Fundamental (high $T$) | Composite of $p_i$ PDWs | $\rho_{q_i} \propto \boldsymbol{d}_{p_i} \cdot \boldsymbol{d}_{p_i}^*$ |
| $h_{1,2}$ | Harmonic-derived | Harmonic of $q_i$ CDWs | $h_1 = q_3 - q_2$<br>$h_2 = q_3 - q_1$ |

# Supplementary Information for

## Evidence of intertwined pair density and charge density wave orders in $UTe_2$

Zhen Zhu[1], Yudi Huang[1], Julian May-Mann[2], Kaiming Liu[1], Zheyu Wu[3], Shanta R. Saha[4], Johnpierre Paglione[4,5], Alexander G. Eaton[3], Andrej Cabala[6], Michal Vališka[6], Eduardo Fradkin[1,7], and Vidya Madhavan[1,5]

8. Department of Physics and Materials Research Laboratory, Grainger College of Engineering, University of Illinois at Urbana-Champaign, Urbana, IL, USA.
9. Department of Physics, Stanford University, Stanford, CA 94305, USA.
10. Cavendish Laboratory, University of Cambridge, Cambridge CB3 0HE, United Kingdom.
11. Maryland Quantum Materials Center, Department of Physics, University of Maryland, College Park, MD, USA.
12. Canadian Institute for Advanced Research, Toronto, Ontario, Canada.
13. Charles University, Faculty of Mathematics and Physics, Department of Condensed Matter Physics, Ke Karlovu 5, Prague 2, 121 16, Czech Republic.
14. Institute for Condensed Matter Theory, University of Illinois, Urbana, IL, USA.

This document includes Supplementary Figures 1-12

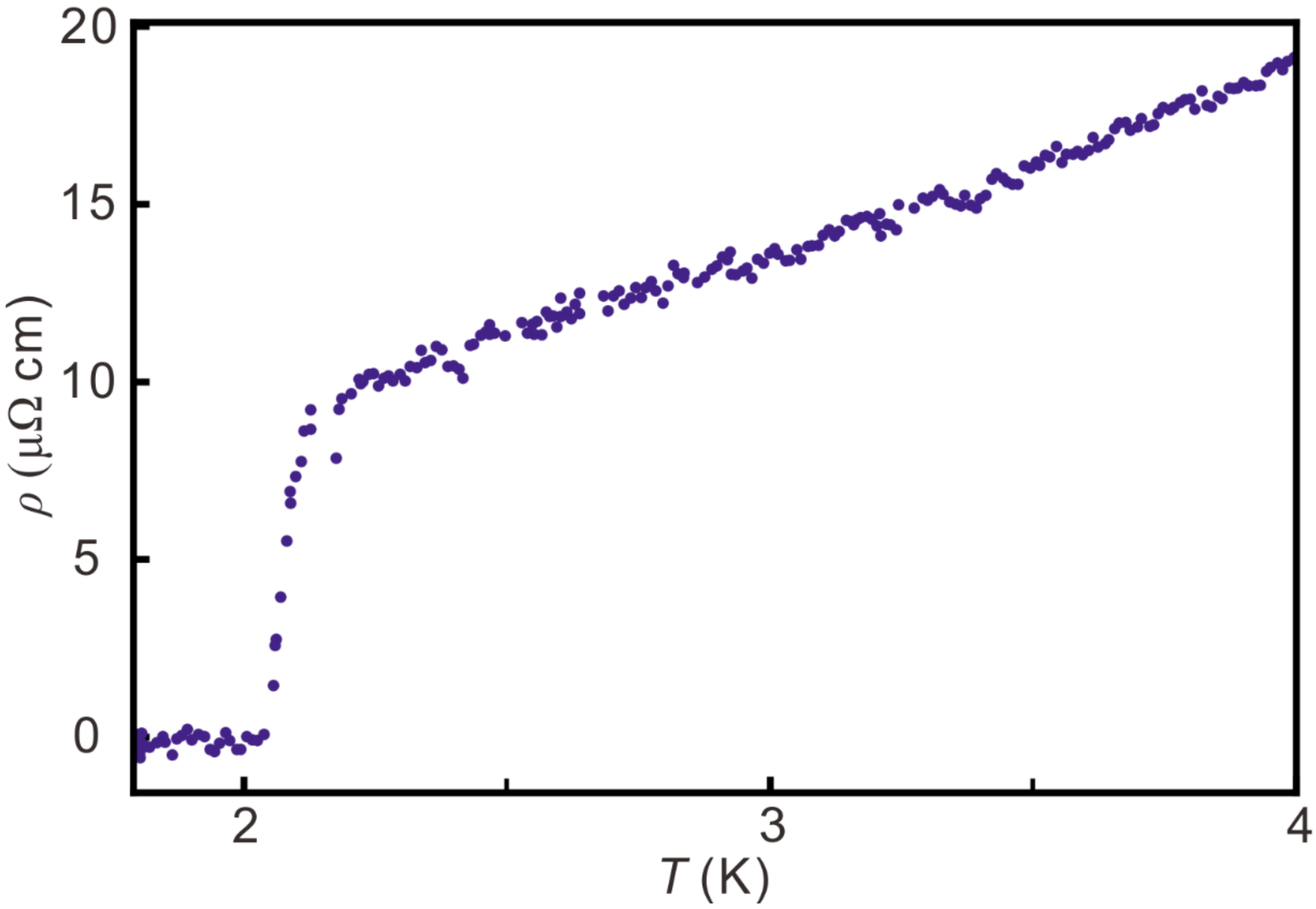

**Supplementary Figure 1| Temperature dependence of resistivity in $UTe_2$ single crystals grown by the molten-salt flux method.** Electrical resistivity ($\rho$) of $UTe_2$ as a function of temperature ($T$) between 1.8 K and 4 K. A sharp drop to zero resistance is observed at ~2.1 K, marking the superconducting transition.

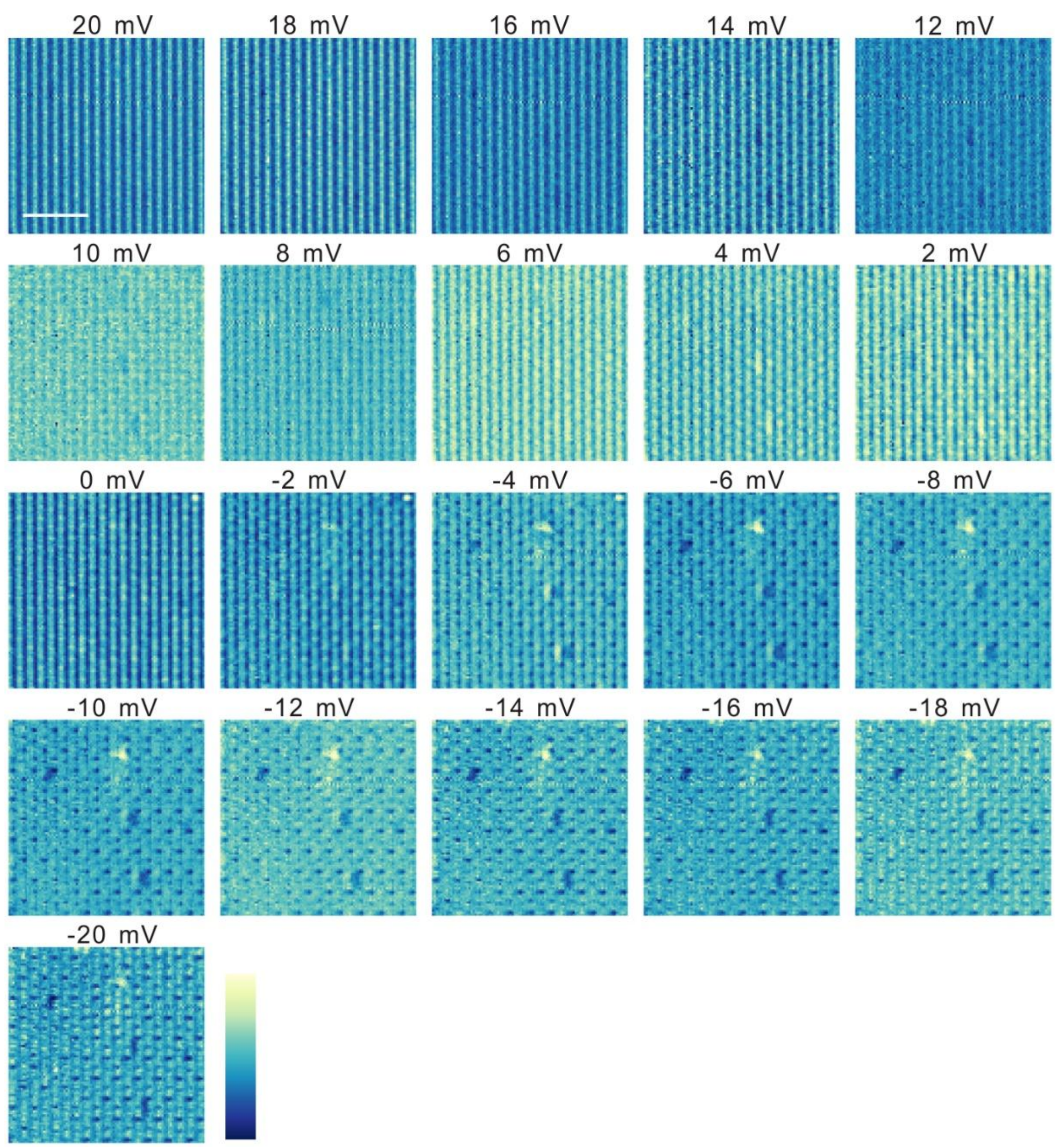

**Supplementary Figure 2| LDOS at 300 mK.** LDOS maps acquired on a defect-free area at the indicated energies. Scale bar, 5 nm. ($V$ = 20 mV, $I$ = 150 pA).

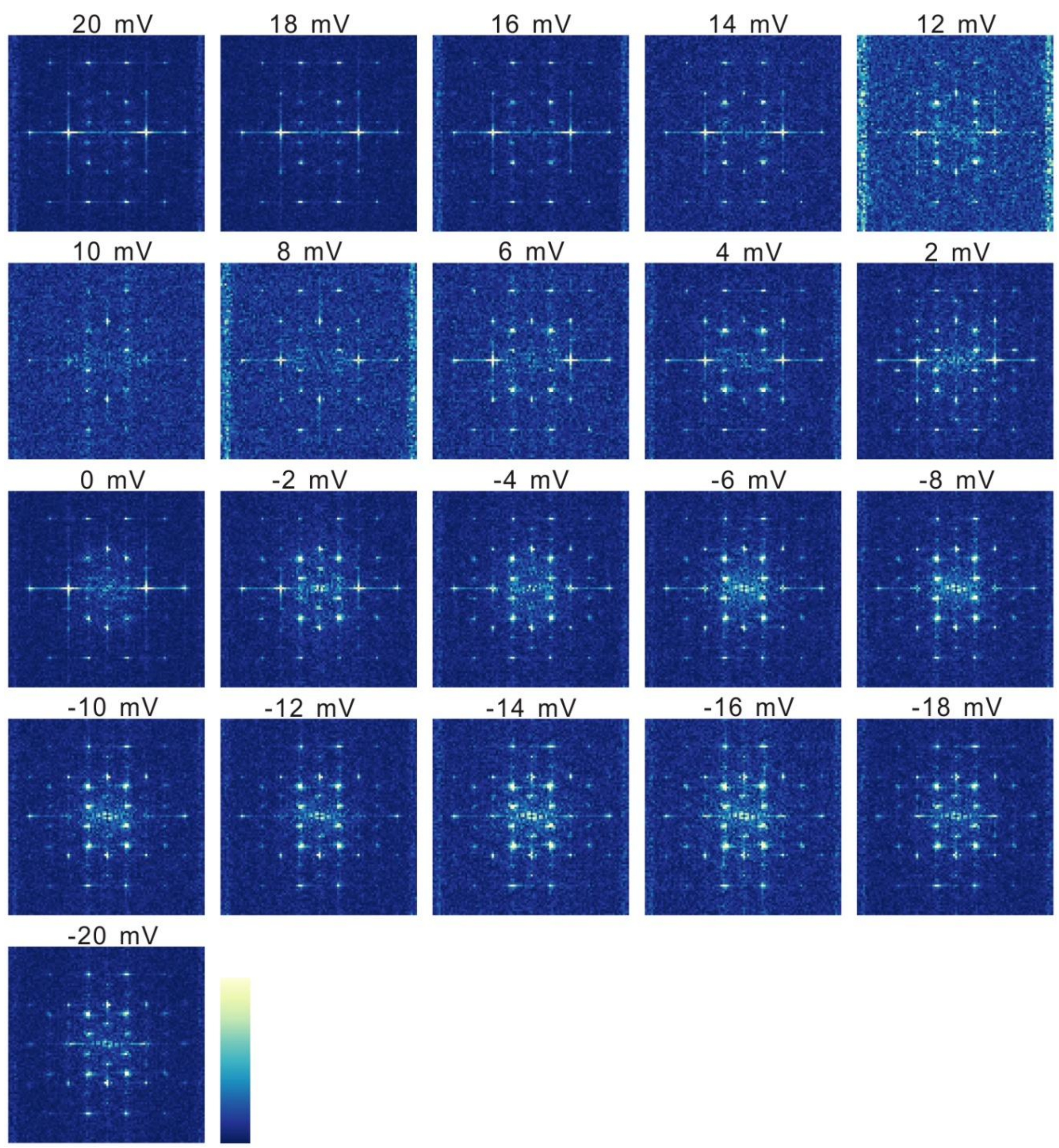

**Supplementary Figure 3| FFT of LDOS at 300 mK.** FFTs of LDOS maps acquired on a defect-free area at the indicated energies.

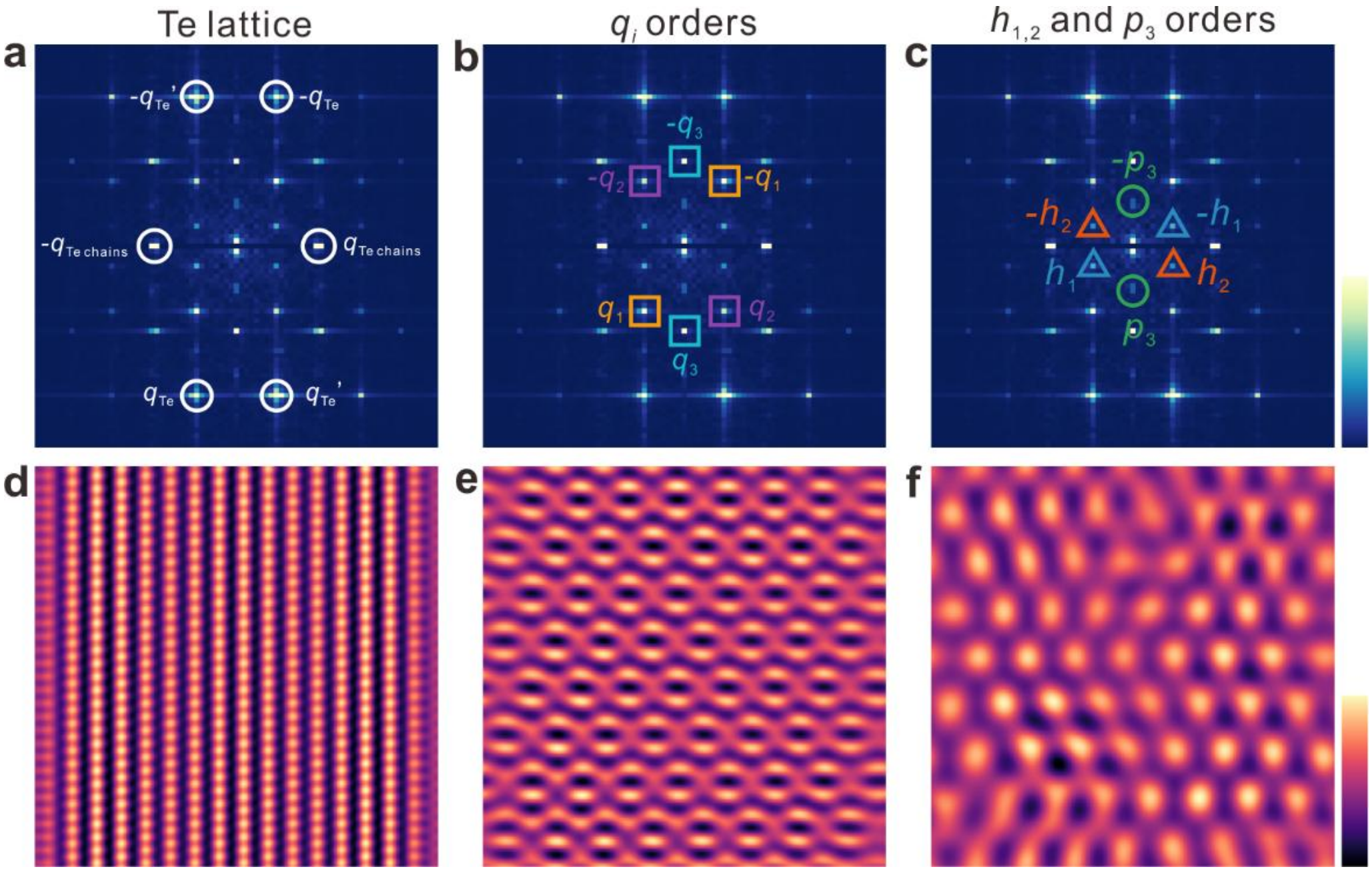

**Supplementary Figure 4| Real-space patterns of the Te lattice, $q_i$ ($i$=1, 2, 3), $h_{1,2}$ and $p_3$ CDW orders at 300 mK under zero magnetic field**. **a-c,** FFTs of STM image in Fig.1c, with Te Bragg peaks, $q_i$, $h_{1,2}$ and $p_3$ peaks marked respectively. **d-f,** Inverse FFTs of the corresponding peaks shown in **a-c**, revealing the real-space modulations associated with each component. Notably, the $h_{1,2}$ and $p_3$ CDW orders form a hexagonal lattice–like periodic modulation.

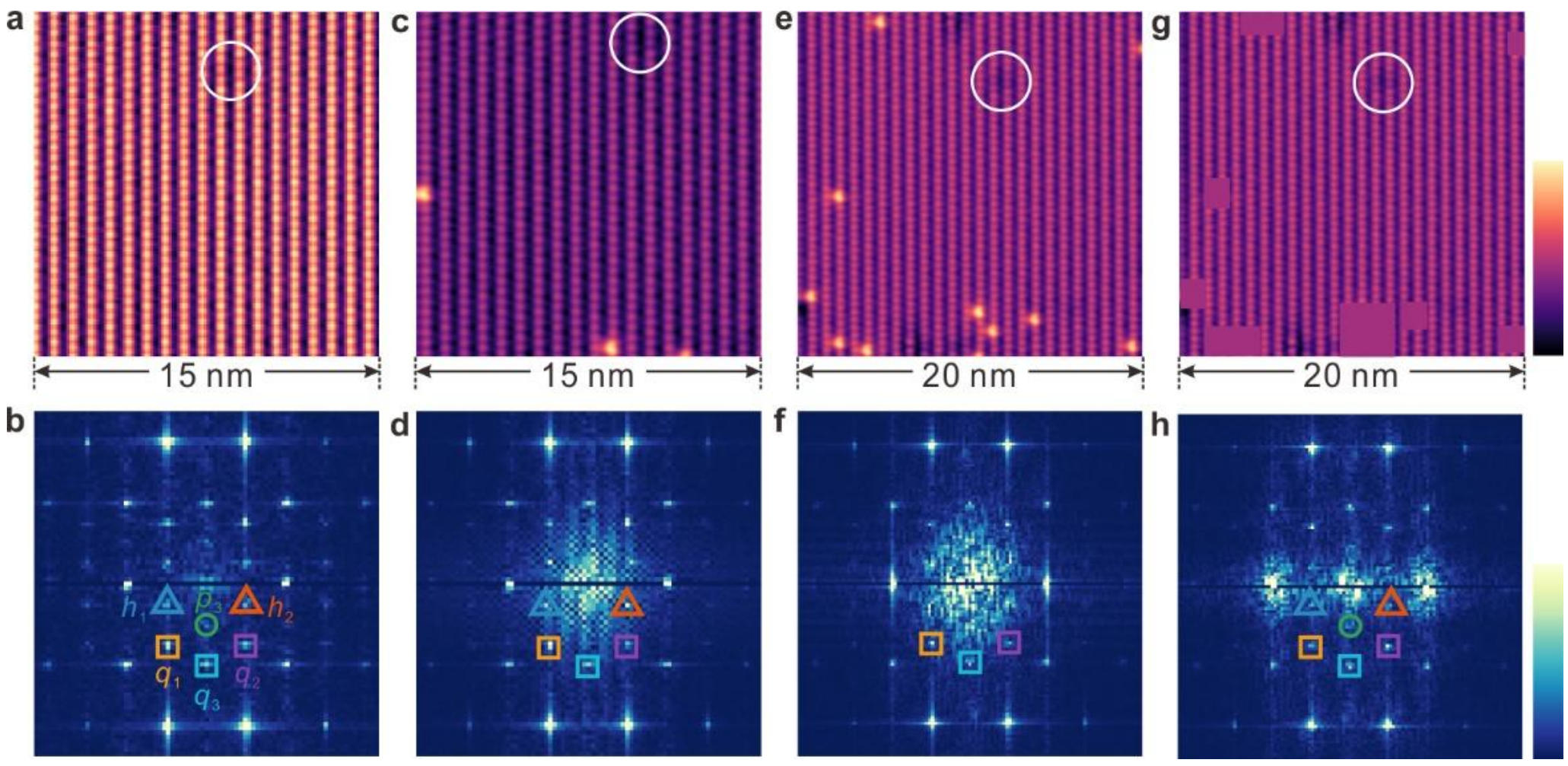

**Supplementary Figure 5| Impact of surface defects on the visibility of the $h_{1,2}$, $p_3$ CDW orders at 300 mK. a, b,** STM topography acquired on a region with no visible top-layer defects, and the corresponding FFT, where $q_i$, $h_{1,2}$, $p_3$ CDW orders are clearly resolved. **c, d,** Topography measured in a nearby region to that shown in **a**, containing a few surface defects, and the corresponding FFT. The $p_3$ peak becomes embedded within defect-induced states. **e, f,** Topography measured over a larger region that includes the area shown in **a** and **c**, with more surface defects present, and the corresponding FFT. The $h_{1,2}$, $p_3$ orders are completely smeared by defect interference and becomes indistinguishable. **g, h,** Topography and FFT after masking the surface defects in **e**, which recover the $h_{1,2}$, $p_3$ CDW signals. White circles mark a subsurface defect that serves as a spatial reference, confirming that the regions shown in **a**, **c**, and **e** were acquired within the same local area. The $q_i$ order remains clearly visible regardless of the presence of surface defects, suggesting that interference from top-layer disorder most likely hindered the detection of the $h_{1,2}$, $p_3$ orders ($T$=300 mK, $V$=10 mV, $I$=100 pA).

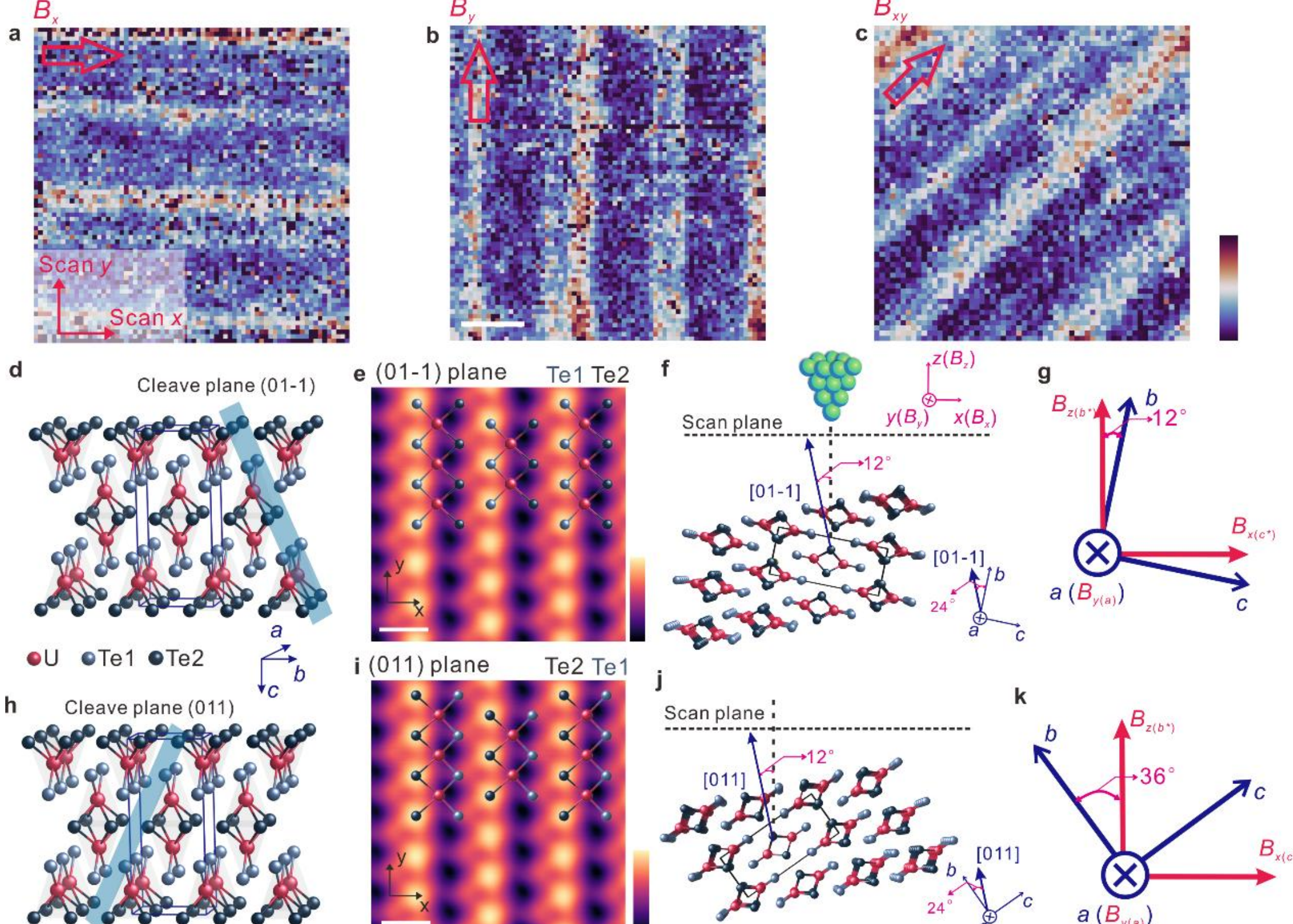

**Supplementary Figure 6| Schematic of magnetic field orientation relative to crystallographic axes. a-c**. In-plane vortices imaged on the surface of 2H-$NbSe_2$ via *dI*/*dV* maps acquired at 0 mV (*T*=300 mK, *V*=3 mV, *I*=200 pA, $V_{mod}$=50 µV). A magnetic field of 0.2 T is applied along the $B_x$, $B_y$, and $B_{xy}$ directions, as indicated by the red arrows. The in-plane field induced vortices exhibit stripe-like structures parallel to the field direction, confirming that the magnetic field axes are aligned with the scanner coordinates. Scale bar, 60 nm. **d**. Schematic of the $UTe_2$ (01-1) cleave plane. **e**. High-resolution topographic image of $UTe_2$ (*V*=50 mV, *I*=100 pA). Based on the crystallographic structure that Te1 atoms lie at a higher absolute height than Te2, the observed dominant chain is attributed to Te1. The overlaid atomic configuration of Te1 (light blue) and Te2 (dark blue) under this assumption corresponds to a (01-1) surface termination. Scale bar, 0.5nm. **f**. Schematic of scan plane and (01-1) plane. The (01-1) plane is tilted by 12° counterclockwise with respect to the scan plane. The [01-1] surface normal forms a constant angle of 24° with the *b* axis. **g**. Orientation of the vector magnetic field directions relative to the crystallographic axes for the (01-1) cleave plane. The field $B_y$ ($B_a$) is aligned with *a*-axis, while $B_x$ ($B_{c^*}$) and $B_z$ ($B_{b^*}$) are tilted counterclockwise by approximately 12° from the *c*- and *b*-axes. **h.** Schematic of the $UTe_2$ (011) cleave plane. **i**. Same as **e**. Based on the calculation that Te2 p orbitals dominate the density of states at the Fermi level, the observed dominant chain is attributed to Te2. The overlaid atomic configuration of Te1 (light blue) and Te2 (dark blue) in this scenario is consistent with a (011) surface termination. **j**. Schematic of scan plane and (011) plane. **k.** Orientation of the vector magnetic field directions relative to the crystallographic axes for the (011) cleave plane. The field $B_y$ ($B_a$) is aligned with *a*-axis, while $B_x$ ($B_{c^*}$) and $B_z$ ($B_{b^*}$) are tilted counterclockwise by approximately 36° from the *c*- and *b*-axes.

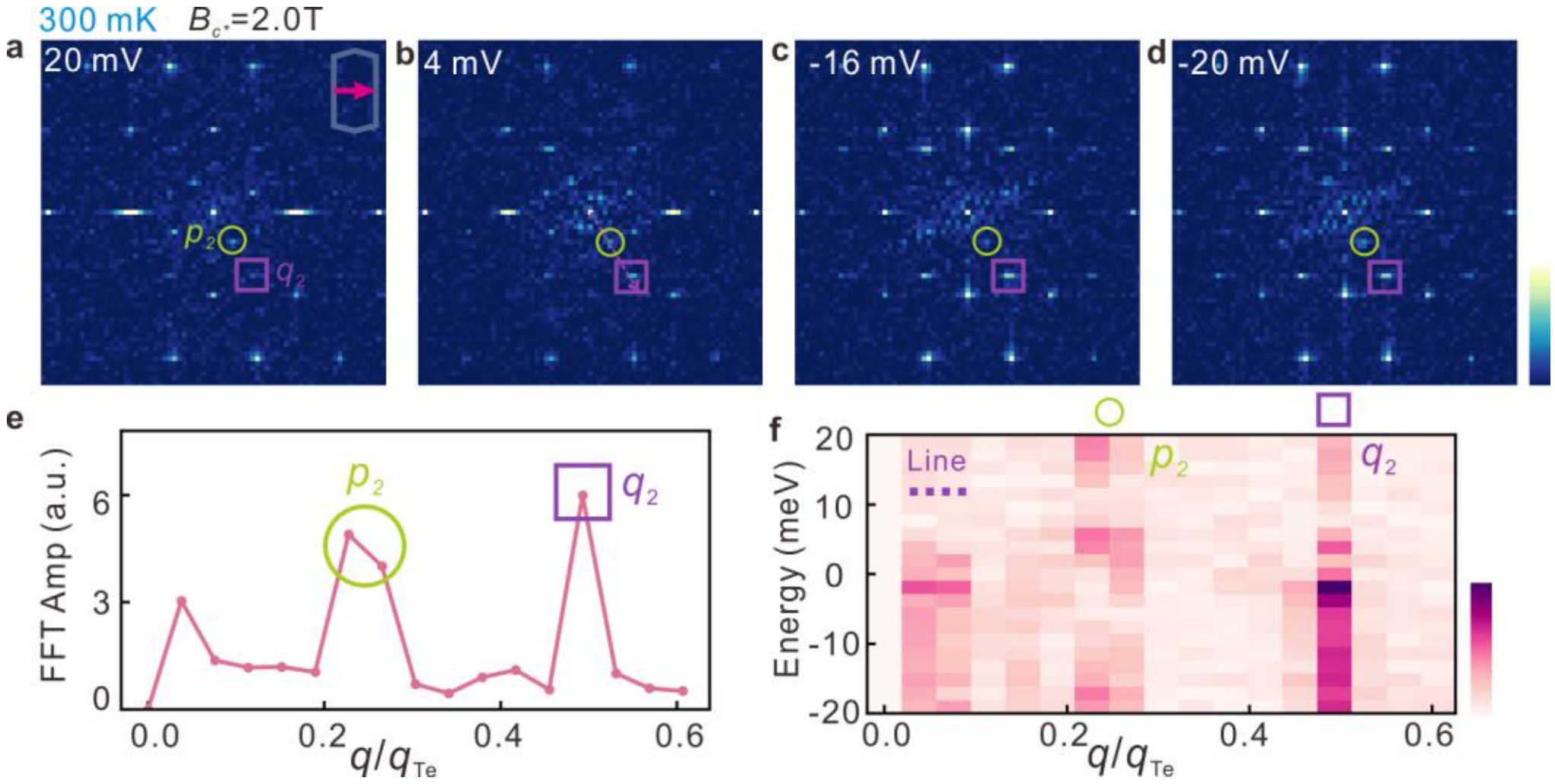

**Supplementary Figure 7| Energy dependence of field-induced $p_2$ CDW at 300 mK under field applied along the $B_{c^*}$ direction. a-d,** FFTs of LDOS maps at the indicated energies under $B_{c^*}$ = 2.0T. The field-induced $p_2$ CDW peak, marked by light green circle, is clearly visible at all energies ($V$=20 mV, $I$=100 pA). **e**, Line cut extracted along the purple line indicated in **b**. The magnitude of $p_2$ is approximately half that of $q_2$. **f**, The intensity map plotted by line cut extracted from FFTs at various energies along the dashed lines in **b.** The magnitude of the $p_2$ wavevectors remains constant with energy, consistent with their identification as CDW signals. The center point is set to zero.

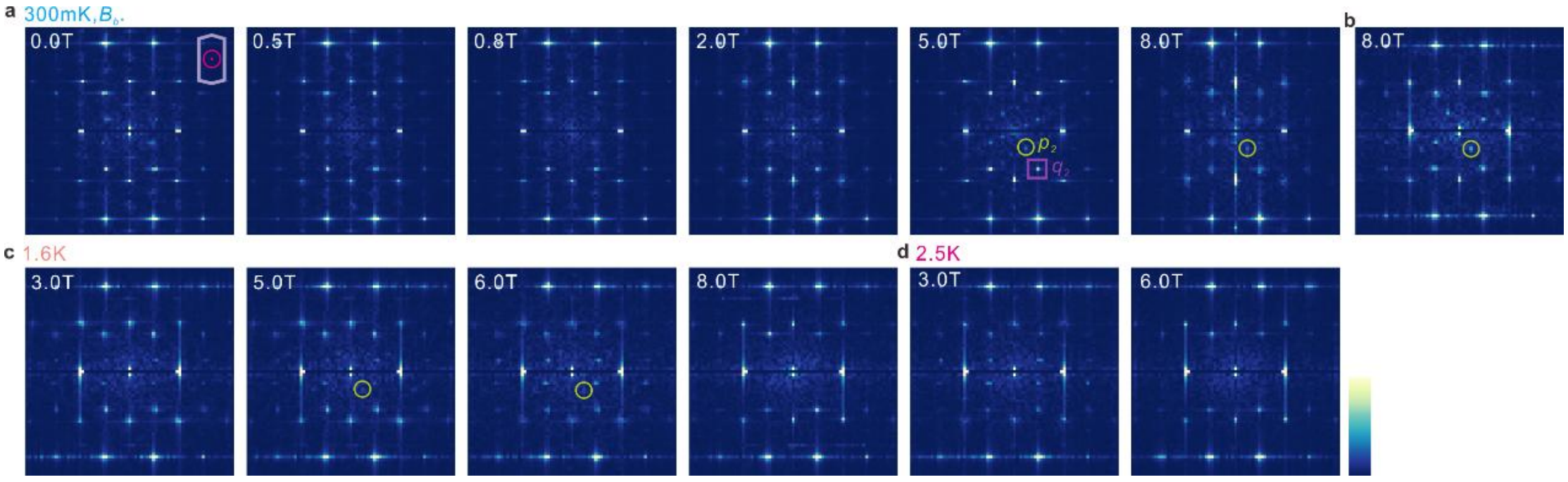

**Supplementary Figure 8| Temperature and $B_{b^*}$-field dependence of the field-induced $p_2$ CDW. a,** FFTs of topographies acquired at 300 mK under magnetic fields applied along $B_{b^*}$ direction. The field-induced $p_2$ peak emerges at 5.0 T and persists at least up to 8 T. **b,** FFT obtained under 8 T at 300mK with the scan frame rotated by 90°, confirming the presence of the $q_i$ CDW order at this field and temperature. The intense central line observed in **a** at 8 T is identified as noise. **c**. FFTs measured at 1.6 K under the same field orientation. The $p_2$ peak remains discernible at 5 T and 6 T, but vanishes at 8 T, accompanied by a complete suppression of the $h_1$ and $h_2$ CDW peaks. **d**, FFTs acquired at 2.5 K, above $T_c$ = 2.1 K. No $p_2$ peak is observed at any field, and the $h_1$ and $h_2$ orders are fully suppressed by 6 T. All measurements were performed in the same area with identical set point conditions ($V$=20 mV, $I$=100 pA), and FFT intensities are normalized by $q_{Te}$ and $q_{Te}'$.

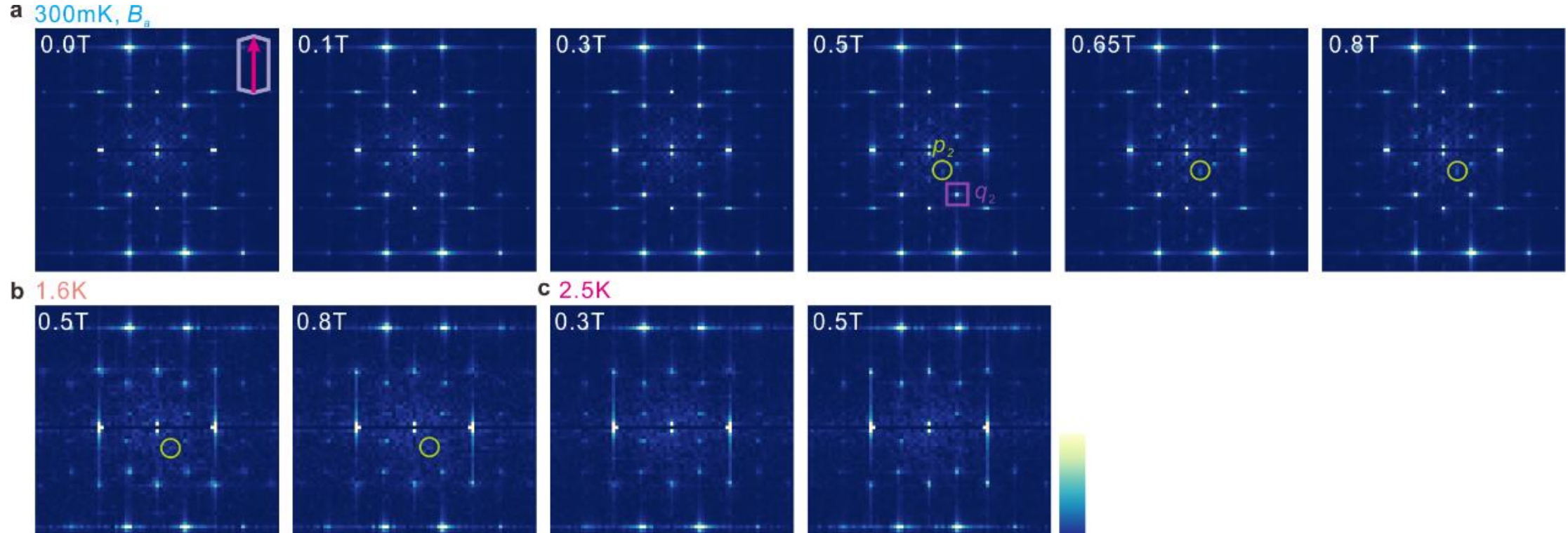

**Supplementary Figure 9| Temperature and $B_a$-field dependence of the field-induced $p_2$ CDW. a,** FFTs of topographies acquired at 300 mK under magnetic fields applied along $B_a$ direction. The field-induced $p_2$ peak emerges at 0.5 T and persists at least up to 0.8 T. **b**. FFTs measured at 1.6 K under the same field orientation. The $p_2$ peak remains discernible at 0.5 T and 0.8 T, while the $h_2$ peak is suppressed at 0.8 T. **c**, FFTs acquired at 2.5 K, above $T_c$ = 2.1 K. No $p_2$ peak is observed at any field, and the $h_2$ peak is suppressed by 0.5 T. All measurements were performed in the same area with identical set point conditions ($V$=20 mV, $I$=100 pA), and FFT intensities are normalized by $q_{Te}$ and $q_{Te}$'.

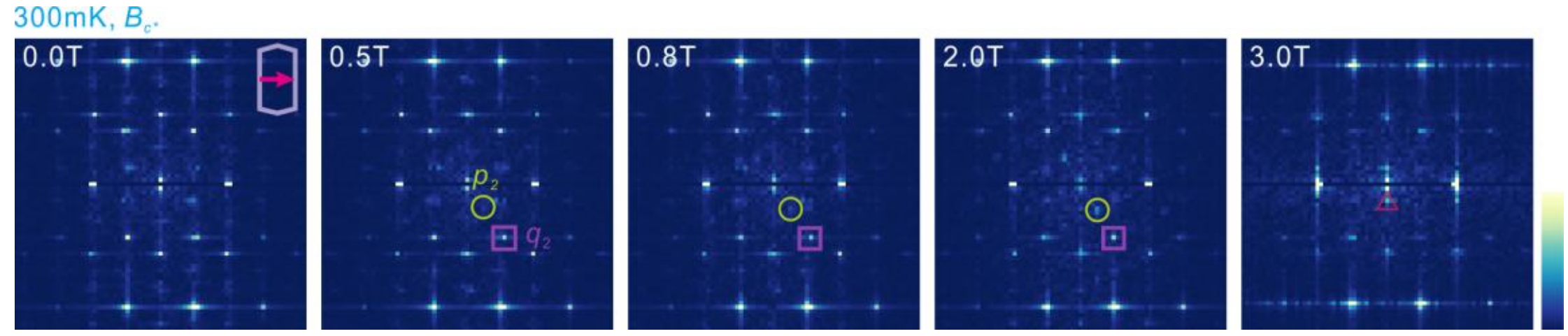

**Supplementary Figure 10| $B_{c*}$-field dependence of the field-induced $p_2$ CDW.** FFTs of topographies acquired at 300 mK under magnetic fields applied along $B_{c*}$ direction. The field-induced $p_2$ peak emerges at 0.5 T and persists at least up to 2.0 T. The $p_2$ signal becomes weaker at 3.0 T, accompanied by the appearance of an additional peak (red triangle) whose wavevector magnitude is half that of $p_3$. All measurements were performed in the same area with identical set point conditions ($V$=20 mV, $I$=100 pA), and FFT intensities are normalized by $q_{\mathrm{Te}}$ and $q_{\mathrm{Te}}'$.

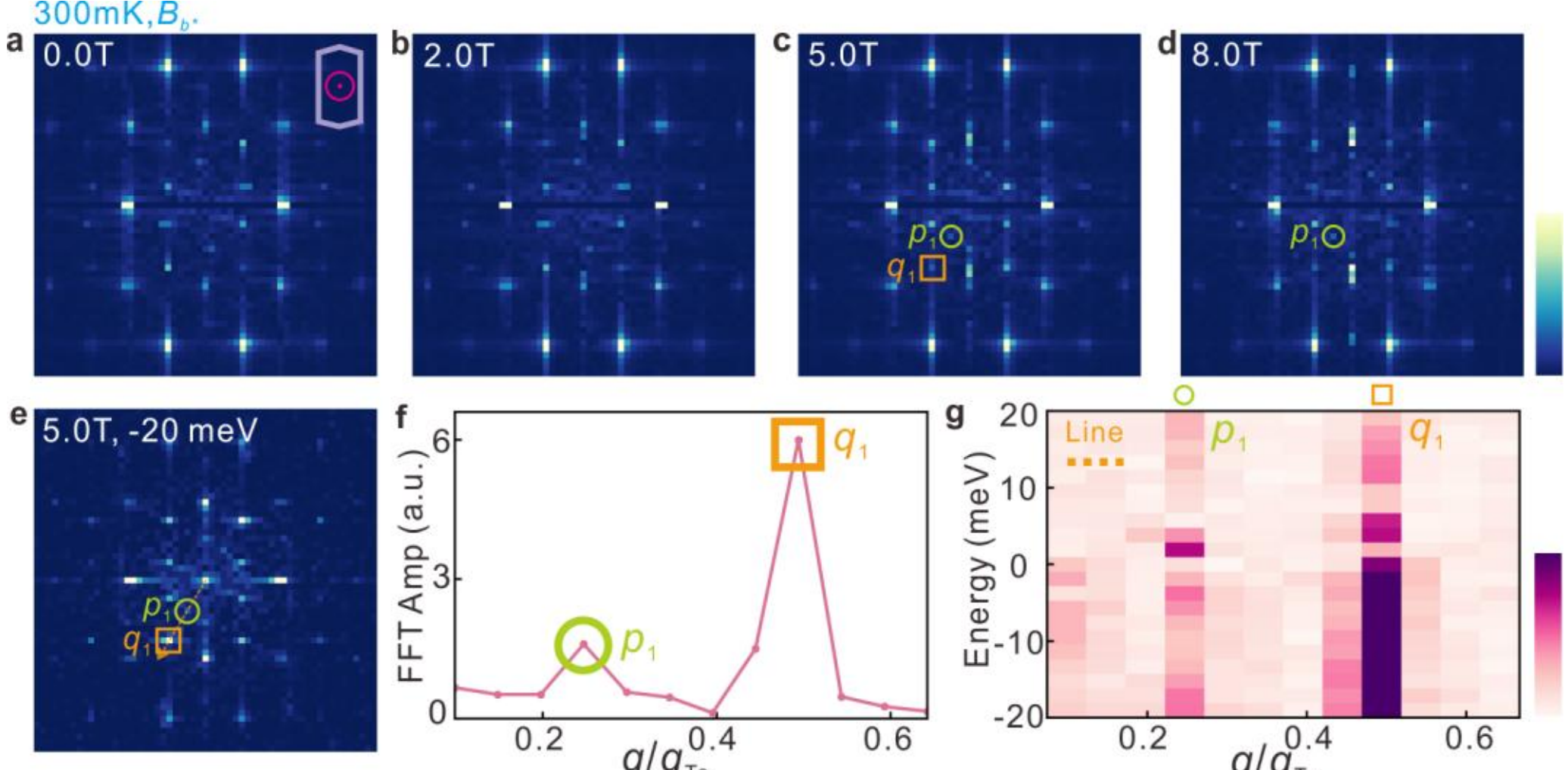

**Supplementary Figure 11| Emergence of $p_1$ CDW in a distinct region under $B_{b^*}$ field. a-d,** FFTs of topographies acquired in the same area at 300 mK under magnetic fields applied along the $B_{b^*}$ direction. This field of view differs from that shown in Supplementary Figures 8–10. A $p_1$ CDW peak, the mirror counterpart of $p_2$, appears at 5 T and 8 T. All measurements were performed with identical set-point conditions ($V$ = 10 mV, $I$ = 100 pA), and FFT intensities are normalized by $q_{Te}$ and $q_{Te}$'. **e,** FFT of the LDOS map acquired at –20 mV and 300 mK under $B_{b^*}$ = 5 T. **f**, Line cut extracted along the yellow line indicated in **e**. The magnitude of $p_1$ is approximately half that of $q_1$. **g,** The intensity map plotted by line cut extracted from FFTs at various energies along the dashed lines in **e.** The magnitude of the $p_1$ wavevectors remains constant with energy, consistent with their identification as CDW signals.

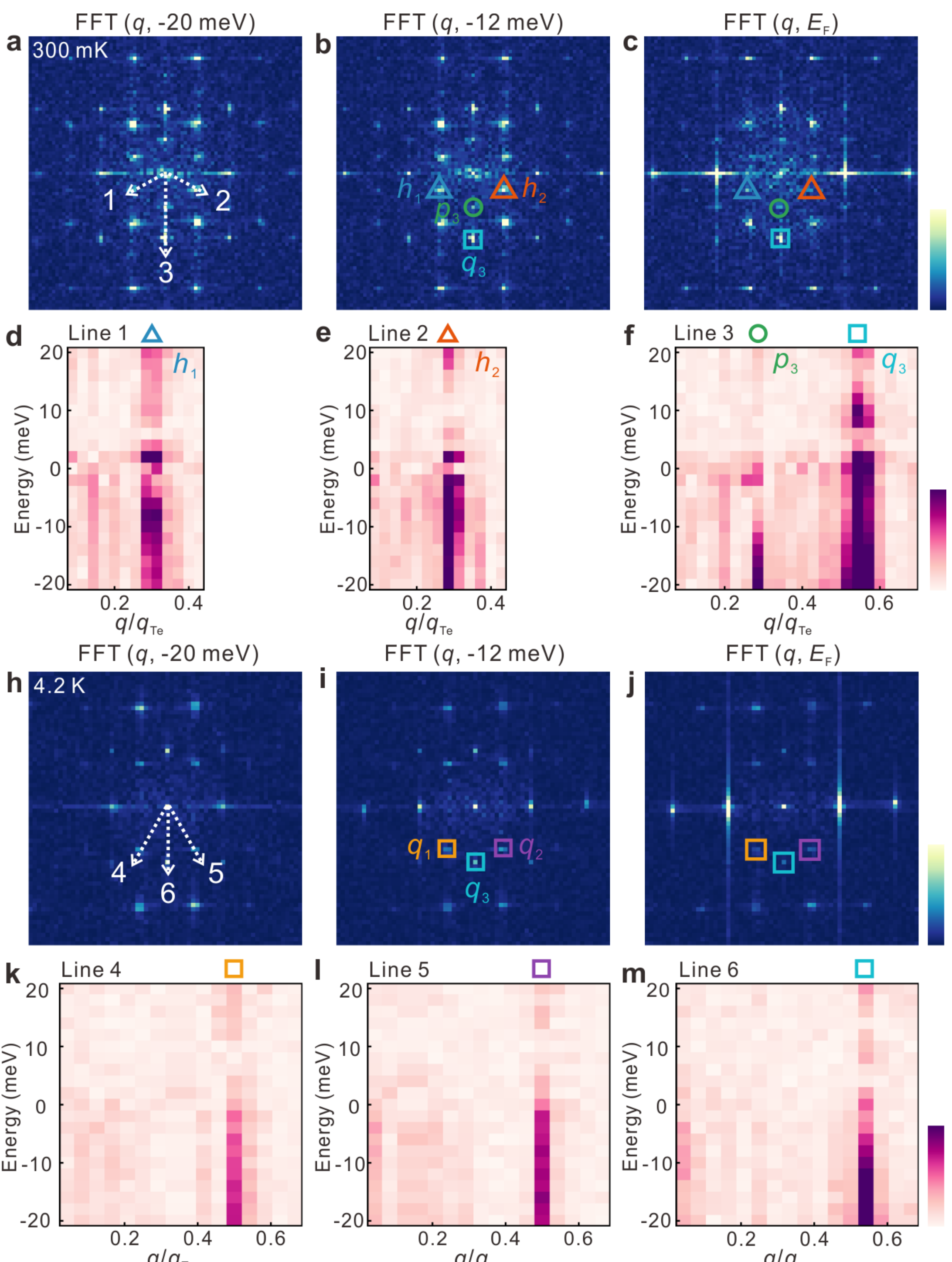

**Supplementary Figure 12| Energy-independent $h_{1,2}$, $p_3$ peaks at 300 mK and $q_{1,2,3}$ peaks at 4.2 K. a-c,** FFTs of LDOS maps acquired at 300 mK for the indicated energies, same as Figs.1 e-g ($V$=20 mV, $I$=150 pA). **d-f,** The intensity maps plotted by line cuts extracted from FFTs at various energies along the dashed lines in **a,** same as Figs.1 i-k. **h-j,** FFTs of LDOS maps acquired at 4.2 K for the indicated energies ($V$=20 mV, $I$=150 pA). **k-m,** The intensity maps plotted by line cuts extracted from FFTs at various energies along the dashed lines in **g.** The magnitude of the $h_{1,2}$, $p_3$ and $q_{1,2,3}$ wavevectors remains constant with energy, consistent with their identification as CDW signals.